\documentclass[conference]{IEEEtran}
\IEEEoverridecommandlockouts
\usepackage{amsmath,amssymb,amsfonts}
\usepackage{algorithmic}
\usepackage{graphicx}
\usepackage{textcomp}
\usepackage{xcolor}
\usepackage{subfigure} 

\newtheorem{theorem}{Theorem}
\newtheorem{lemma}{Lemma}
\newtheorem{remark}{Remark}
\newtheorem{definition}{Definition}

\usepackage{hyperref}

\def\BibTeX{{\rm B\kern-.05em{\sc i\kern-.025em b}\kern-.08em
    T\kern-.1667em\lower.7ex\hbox{E}\kern-.125emX}}
\begin{document}

\title{Faster Data-access in Large-scale Systems: Network-scale Latency Analysis under General Service-time Distributions\\
\vspace{-3mm}
}
\author{\IEEEauthorblockN{Avishek Ghosh}
\IEEEauthorblockA{\textit{Department of EECS} \\
\textit{University of California, Berkeley, USA}\\
avishek\_ghosh@berkely.edu}
\vspace{-10mm}
\and
\IEEEauthorblockN{Kannan Ramchandran}
\IEEEauthorblockA{\textit{Department of EECS} \\
\textit{University of California, Berkeley, USA}\\
kannanr@eecs.berkeley.edu}
\vspace{-10mm}}

\maketitle

\begin{abstract}
In cloud storage systems with a large number of servers, files (e.g., videos, movies) are typically not stored in  single servers. Instead, they are split,  replicated (to ensure reliability in case of server malfunction) and stored in different servers. We analyze the mean latency of such a split-and-replicate cloud storage system under general sub-exponential service time distribution, which encapsulates most of the practical heavy-tailed distributions. We present a novel scheduling scheme that utilizes the load-balancing policy of the \textit{power of $d$ $(\geq 2)$} choices. Exploiting the \textit{double exponential queue length} property of this policy (\cite{bramson_sampling}), we obtain tight upper bounds on mean latency. An alternative to split-and-replicate is to use erasure-codes, and recently, it has been observed that they can reduce latency in data access (see \cite{longbo_delay} for details). We argue that under high redundancy (integer redundancy factor strictly greater than or equal to 2) regime, the mean latency of a coded system is upper bounded by that of a split-and-replicate system (with same replication factor) and the gap between these two is small. For example, when specialized to an exponential service time distribution, our formulation recovers the result of \cite{srikant_mean}, (which uses erasure codes) upto a constant factor. We also validate this claim numerically under different service distributions such as exponential, shift plus exponential and the heavy-tailed Weibull distribution and compare the mean latency to that of an unsplit-replicated system. We observe that the coded system outperforms the unsplit-replication system by at least $20\%$ for all three distributions and all possible arrival request rates. Furthermore, we consider the mean latency for an erasure coded system with low  redundancy (fractional redundancy factor between 1 and 2), a scenario which is more pragmatic, given the storage constraints (\cite{rashmi_thesis}). However under this regime, we restrict ourselves to the special case of exponential service time distribution and use the randomized load balancing policy namely \textit{batch-sampling}. We obtain an upper bound on mean delay that depends on the order statistics of the queue lengths, which, we further smooth out via a discrete to continuous approximation. 
\end{abstract}

\begin{IEEEkeywords}
Cloud Storage Systems, Latency, Erasure Codes
\end{IEEEkeywords}

\vspace{-5mm}
\section{Introduction}
\begin{figure}[t!]
    \centering
    \includegraphics[width=0.5\textwidth]{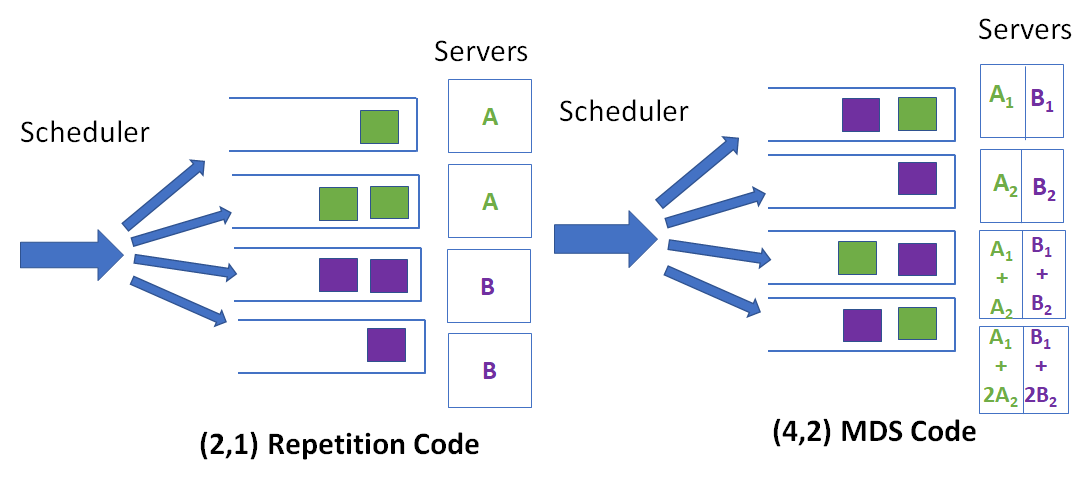}
    \vspace{-5mm}
    \caption{An example of a data center with 4 servers with naive-replication (left) and coded (right) with a $(4,2)$ MDS code. There are two different types of files $A$ (green) and $B$ (purple). Upon a file-request arrival for a particular file, the scheduler advances the request to that subset of the servers containing the entire (for naive-replication) or portions (for coding) of the file.
     }
    \label{fig:example}
    \vspace{-3mm}
\end{figure}

In large data centers, multitudes of servers are deployed to serve user requests. We consider a model where each server is associated with a queue of infinite capacity and the service requests are queued. The service principle in the servers is ``work conservative'', i.e., the servers are busy when the queue lengths are non-zero. We want the entire system to be robust to individual server failures. A naive way to achieve this is via replicating the entire file and store in different servers (we will call this scheme as ``naive replication''). A possible alternative is to split the file into chunks, replicate each chunk, and store them in different data servers (i.e., split-and replication scheme). Furthermore, one can also use  Maximum Distance Separable (MDS) codes \footnote{An $(n,k)$ MDS code is defined in the following way: if we have $k$ message symbols and $n$ encoded symbols and at most $n-k$ symbols are erasured we can still retrieve the $k$ original message symbols. } to achieve this. There has been a significant amount of research work that focuses on the fact that if Maximum Distance Separable (MDS) codes are used instead of naive-replication, we gain in terms of storage and repair cost. Classically, MDS codes in data centers are used for precisely this purpose \footnote{Other than  classical applications, MDS codes are useful for optimal repair bandwidth (\cite{wang-repair}) and for regeneration codes for distributed storage repair (\cite{suh-reg})}. 

Another major concern in data centers is to provide data access with minimum latency, and it turns out that both split-and-replicate and MDS codes can be beneficial in this aspect (\cite{longbo_delay}) as compared to the naive replication. We will illustrate the effectiveness of codes via a simple example. Consider a data center with $4$ servers and having $2$ files $A$ and $B$. In a replication scenario, assume each file is replicated twice and stored in 4 servers as shown in Figure~\ref{fig:example}. In this system, if the requested file is file $A$, the scheduler forks the request to either server 1 and 2. Also, for now, assume that the rate of request is very low, i.e., the queue lengths are almost zero at the servers. If we assume that the service time for each server is denoted by the random variable $X_r$, the average time to read file $A$ will be $\mathbb{E}(X_r)$.

 Now, let us analyze the erasure coded system. We now consider a $(4,2)$ MDS code, where each file is chunked into 2 parts $A_1$ , $A_2$ and $B_1$, $B_2$, and 2 other parity chunks, $A_1+A_2$ and $A_1+2A_2$ (similarly for $B$) are used (Refer Figure~\ref{fig:example}). From the property of MDS codes, it is sufficient to read any 2 out of this 4 chunks to recover the entire file. Since we are assuming no queuing effects, we can, without loss of generality assume that the request is served by servers $1$ and $2$, and hence we observe that, the time to read any entire file (file $A$ for instance) will be $\mathbb{E}\lbrace \max(X_c^1,X_c^2) \rbrace $, where the service time of the $i$-th server is denoted by $X_c^i$, and we assume that $X_c^i$, for $i=\lbrace 1,2 \rbrace$ are independent and identically distributed.  Since each file is chunked into 2 halves, for large payloads, we can expect the average reading time will be halved in this scenario, i.e., $\mathbb{E}(X_c^i) \approx\frac{1}{2}\mathbb{E}(X_r)$. 

The gain (over naive replicated system) in using MDS codes is characterized by the difference in expected download time of any file under naive-replication and erasure coded system with identical storage constraints. In this example, the gain is $G = \mathbb{E}(X_r)-\mathbb{E}\lbrace \max(X_c^1,X_c^2) \rbrace$. As an example, we first choose  $X_r \sim \exp(1)$ and $X_c^i \sim \exp(2)$. A simple calculation gives, $G=0.25$, which is positive, and hence it is advantageous to use coding over naive-replication. The service time distribution in data centers typically is not exponential  as observed in \cite{ec_cache_rashmi}. It is observed \cite{lee_distributed} that ``shift plus exponential'' (see \cite{lee_distributed} for experimental validation) distribution along with heavy tailed distributions (like Pareto distribution, Weibull distribution) approximately captures the service discipline of servers in a large server system (refer \cite{ec_cache_rashmi} for details). In this paper, we will provide theoretical analysis for the above mentioned distributions.

We now choose the ``shift + exponential'' distribution. Intuitively, this is an accurate model for data centers because servers usually have an overhead (which may result from the scheduling discipline) to serve before the actual work-load. Under this discipline, $X_r = c+Y_r$, where $c>0$ is a small constant and $Y_r\sim \exp(1)$. Similarly, $X_c^i = \frac{c}{2}+ Y_c^i$, where $Y_c^i \sim \exp(2)$. The shift $c$ can be seen as a response time characteristic of the server. For $c=0.2$, we get $G = 0.35 $. Note that, in this example, we have assumed that a read request implies reading the entire segment of the file (the entire file for naive-replication system and an entire chunk for erasure coded system) and no partial reads are allowed. The intuitive gain observed in these two examples motivate us to investigate the scenario where the arrival request rate is non zero, i.e., when the queuing effect cannot be ignored. In this work, we rigorously show that even with queuing effects, both split-and-replicated and  coded systems outperform naive replication-based systems.

Throughout the paper, we provide sharp analysis for the mean latency of data centers. Another useful measure is to compute the tail-latency (\cite{ec_cache_rashmi}). We provide a brief analysis of tail latency for exponentially distributed servers.

\subsection{Related Work and Our Contribution}

Analysis of various load balancing schemes for large data centers under Poisson arrival and memoryless service is a classical problem (refer to \cite{mitzenmacher_power_of_two} and the references therein). \cite{bramson_sampling} generalized the setting for the case of general service time distribution. However, these works generally do not focus on techniques to reduce latency (such as split-and-replicate or coding). In this paper, we combine the load balancing principles with latency reducing techniques.

Erasure coded systems have been studied extensively in the coding theory literature. The delay improvement aspect of erasure coding is first shown in \cite{longbo_delay}. The delay-storage tradeoff for an erasure coded system is explained in \cite{joshi_fork} with the fork-join approach. \cite{joshi_fork} provides bounds on $(n,k)$ fork-join systems, which exploits flooding and assumes zero cancellation cost. \cite{joint_delay_storage} targets the problem from the point of view of joint optimization of storage cost and delay incurred. Very recently, a mean field based analysis of a high redundancy erasure coded system over naive-replication is reported in \cite{srikant_mean}, with exponential service time distribution. In \cite{srikant_mean} the entire system is modeled as a Markov chain, and with a careful characterization of the transition probabilities, an upper bound on mean latency and the gain over replication based systems is obtained. We now summarize the main contributions of the paper:

\vspace{-1mm}
\begin{itemize}
\item We analyze both split-and-replicate and erasure coded data centers, under large system limit. Most prior works analyzed latency performance from a ``local" individual job basis.  In contrast, in this paper, we do a ``global" network-scale latency analysis. We use the load balancing policy of \textit{power of d $(\geq 2)$} choices, which enables us to work with general service time distribution.

\item We argue that, the mean latency of a erasure coded system with high redundancy is upper bounded by that of a split-and-replicate system with same replication factor, and the bound is reasonably tight. \cite{joshi_fork} also deals with general service time distribution, but the results are applicable only for ``local'' fork-join systems. \cite{srikant_mean} deals with the large system, but the analysis presented there holds only for exponential service distribution, where, our framework can accommodate servers with general sub-exponential service time distribution and tailoring to the exponential service distribution, we recover their results upto a constant factor.

\item  One shortcoming of the high redundancy coded system is that it is often prohibitive in many practical scenarios, even with the minimum redundancy factor of $2$. In practice, typically erasure codes with redundancy $1.2-1.5$ are used (for example Facebook data center uses an $(14,10)$ MDS code \cite{rashmi_thesis}). We provide an analysis of the mean file-access delay even when the redundancy is low (i.e., between $1$ and $2$), using \textit{batch sampling}. However, in this low redundancy regime, our analysis only holds for exponential service distribution.
\end{itemize} 

%
%
 \vspace{-2mm}
 \subsection{System Model}
\label{sec:system_model}

We assume a cloud storage system with $L$ servers, and each of them stores a large number of different files. Each file is stored in $n$ servers.  Also assume that there are a total of $I=\Theta(L\log L)$ files, and all files are being requested uniformly at random.  The arrival process is modeled as a Poisson process with rate $L\lambda$, where $\lambda \in (0,1)$. Also each arrival requests a file uniformly at random out of $I$ files. 


We use a split-and-replicate based system to store files. Each file is chunked into $k$ parts, and each part is replicated $d (\in \mathbb{Z}_+)$ times, and stored to $n=dk$ different servers. Alternatively, we also use MDS code to encode files. For example, using an $(n,k)$ MDS code, each file is divided into $k$ parts and encoded into $n$ chunks and stored in $n$ distinct servers in the data center. We assume that the service times of different servers are independent and identically distributed (\textit{iid}) with, mean $\frac{1}{k}$. Alternatively, for a naive-replication based system, the service time, $X_r$, satisfies, $\mathbb{E}(X_r)=1$.

\subsection{$k$-split Scheduling Policy}

The scheduling policies we employ are load balancing policies. For the split-and-replicate system, upon the read (download) request of a file, the scheduler forks the request to $k$ servers out of $n=dk$ servers. The $k$ servers are chosen as follows: we divide the $dk$ servers into $k$ batches of $d$ servers each, where each batch contains replicas of the same chunk of the file, and read the server with the smallest queue length from the each batch. Surely, reading $k$ chunks is equivalent to reading the entire file. For the erasure coded system, we do have a little flexibility, as reading any $k$ out of $n$ servers is sufficient to read the entire file. Here, the scheduler sends the request to those $k$ servers having minimum queue lengths.

For a particular file, the scheduler forwards the request to a deterministic set of $k$ servers. But, we assume that all the files are being requested uniformly at random. Therefore, from this point of view of the scheduler, upon each request, it randomly sends it to a set of $k$ distinct servers. This observation has led us to leverage some properties of randomized load balancing schemes with large system limit.

\begin{figure}[t!]
\centering
{\includegraphics[height=1.6in,width=2.9in]{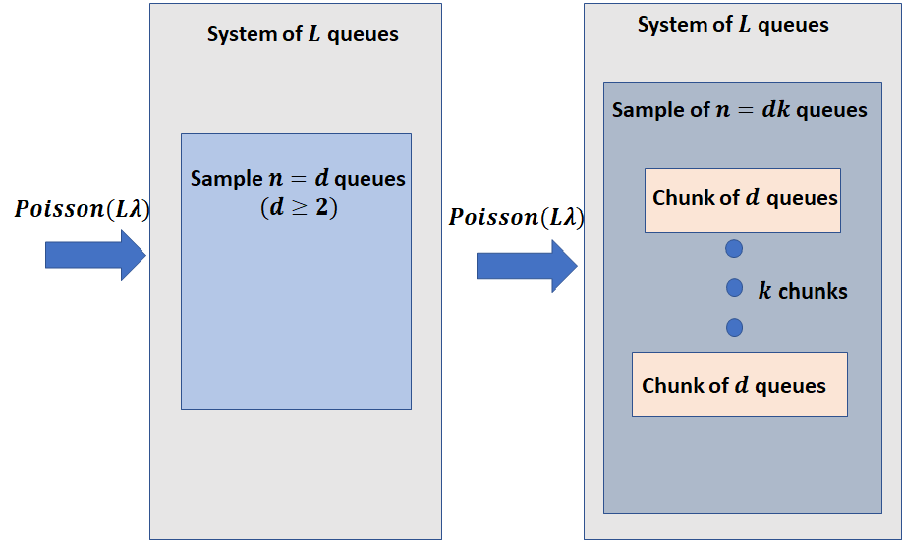}} 
\vspace{-2mm}
 \caption{Scheduling policy for naive replicated system (left)($k=1$) and a split-and-replicate or an erasure coded system (right). In the erasure coded-system, we use the $k$-split policy.}
\label{fig:rep_vs_era} 
\vspace{-3mm}
\end{figure}
 
%
%
%
%
%

\vspace{-1mm}
\section{Latency of Split-and-Replicate System}
\label{sec:method}

 In this section we will analyze the performance of a split-and-replicate system , where a file is chunked in $k$ parts, with each replicated $d (\geq 2) $ times, and stored in $n=dk$ servers. To analyze the performance under this scenario, we will exploit a randomized load balancing policy, namely the power of $d \geq 2$ choices (refer \cite{bramson_sampling}, \cite{mitzenmacher_power_of_two} for details).  

\subsection{Power of $d$ $(\geq 2)$ load balancing policy}
 Assume the request arrival process is Poisson with rate $L \lambda$ to a bank of $L$ servers where $\lambda \in (0,1)$. Upon a file-request, the scheduler randomly samples $d \geq 2$ queues and join the queue with the shortest queue length. In \cite{bramson_sampling}, this scheduling policy is analyzed under different classes of service time distribution (as given in Theorem~\ref{thm:gen_1} and \ref{thm:gen_2}) and it is proved that under the steady state all the queues in the system will be identically distributed with a queue length distribution that decays double exponentially when $L$ is large. In the limit $L \rightarrow \infty$, assume that, $p_m$ denotes the probability that a queue, in steady state, has a minimum queue length of $m$. Here we present the results of \cite{bramson_sampling}:

\begin{theorem}[Class I]
\label{thm:gen_1}
Let $X$ be the service time and suppose we have $\mathbb{E}(\exp(\theta X)) < \infty$ for some $\theta > 0$, then for all $d \geq 2$, we have
$\log_d \log \frac{1}{p_m} = (1+o(1))m$.
\end{theorem}
\vspace{1mm}
\begin{theorem}[Class II]
\label{thm:gen_2}
Let $X$ be the service time and suppose we have $\mathbb{P}(X > x)=\Theta(x^{-\beta})$ for some $\beta > \beta^*$, with $\beta^*=\frac{d}{d-1}$, i.e., the tails of the service time decays faster than $\beta^*$ then for all $d \geq 2$,
$ \log_d \log \frac{1}{p_m} = (1+o(1))m$.
\end{theorem}
Using this, we will now analyze the split-and-replicate system. Let the average reading time of such a system be $\bar{W}_s(n,k)$.

\vspace{-1mm}
\subsection{Mean Latency}

We will analyze the mean delay of a split-and-replication system with sub-exponential service time belonging to Class I and II. Recall that for such a system, the file reading will be complete when reading from all $k$ queues will be complete. We use the $k$-splitting scheduling scheme to choose the servers. Suppose the queue length of these $k$ queues are $\hat{Q}_i(n,k)\, (i=1,\ldots,k)$. So, the job experiences an average delay of,
\vspace{-3mm}
\begin{equation}
\bar{W}_s(n,k)=\mathbb{E}[ \max_{i=1,2,\ldots,k} ( \sum_{j=1}^{\hat{Q}_i(n,k)} X_c^{(j,i)}  +R_i) ] \label{eqn:exact_delay}
\vspace{-2mm}
\end{equation}
where the service time for job $j$ of queue $i$ is $X_c^{(j,i)}$, and $R_i$ denotes the residual service time in queue $i$.

 Also for all $i,j$, $X_c^{(j,i)}$ are iid with mean $\frac{1}{k}$ (refer to the system model in Section~\ref{sec:system_model}). We also assume that $R_i$ are iid. If the service times are exponential, i.e., $X_c^{(j,i)} \sim \exp(k)$, we can use the memory-less property to verify $R_i$ and $X_c^{(j,i)}$ are identically distributed for all $i,j$. The mean delay will be,
 \vspace{-2mm}
\begin{equation}
\bar{W}_s(n,k)=\mathbb{E}[ \max_{i=1,2,\ldots,k} \sum_{j=1}^{\hat{Q}_i(n,k)+1} X_c^{(j,i)}  )] \label{eqn:exact_delay_exp}
\vspace{-2mm}
\end{equation}

\section{Analysis of Mean Latency with Sub-Exponential Service time}
 \label{sec:general_dist}
 
  In this section, we analyze the $(n=dk,k)$ split-and-replicate system with sub-exponentially distributed servers and derive upper bound on mean and tail latency under general sub-gaussian service distribution. Without loss of generality, throughout the section, we assume $d=2$. Owing to space constraints, all the proofs in this section is deferred to the Appendix~\ref{appendix:int}.

 Recall from Section~\ref{sec:system_model} that, we will sample $k$ batch of $d(=2)$ queues (with each batch having identical replicas of a chunk) in parallel and fork the input request to the servers with smallest queue lengths among the batches. Since we are analyzing the system under the large system limit, $i.e., L \rightarrow \infty$, we can assume that under the $k$-split policy the entire system decomposes to $k$ identical such systems with arrival rate of $\frac{\lambda L}{k}$ (owing to independent Poisson splitting). Since the number of servers for each system is $\frac{L}{k}$, and $L \rightarrow \infty$, we can analyze the subsystems under large system limit with arrival rate $\lambda L'$  and number of servers, $L'$ (where $L'=\frac{L}{k} \rightarrow \infty$). 
 
  We assume $X_c^{(j,i)}$ to be sub-exponential. This is a valid assumption since all heavy tailed distribution like Weibull, Pareto distribution follow sub-exponential property (\cite{goldie_subexp}). We now define a sub-exponential random variable.
  \vspace{1mm}
  \begin{definition}
A random variable $Y$ with $\mathbb{E}{Y}=\mu$ is called $(\tau^2,b)$-sub-exponential if $
\mathbb{E}{e^{\lambda(X-\mu)}} \leq e^{\frac{1}{2}^2\lambda^2 \tau^2},~\forall~ | \lambda | <\frac{1}{b}$.
\end{definition}
\vspace{1mm}
As stated in Equation~\ref{eqn:exact_delay}, under general service time distribution, the statistics of the residual service time distribution will have an effect on the mean delay. Also, we can compute the distribution of $\hat{Q}_i(n,k)$ using the $k$-split scheduling policy. From \cite{bramson_sampling}, the distribution of $\hat{Q}_i(n,k)$ will be double exponential: $
  \mathbb{P}(\hat{Q}_i(n,k) \geq r) \leq c_u (\frac{\lambda}{k})^{2^r}
  $ where $c_u = 1+o(1)$. For simplification, we assume $c_u=1$ for analysis. For a general sub-exponential service time, with parameters $(\tau^2,b)$, we have the following result:
  
  \begin{theorem}
\label{thm:result_1_gen}
Under the large system limit ($L \rightarrow \infty$), with $\mathbb{E}( X_c^{(j,i)}) = \frac{1}{k}$, and $X_c^{(j,i)}$ sub-exponential with $(\tau^2,b)$ satisfying, $\tau=\mathcal{O}(\frac{1}{k^{\beta_1}}), b= \mathcal{O}(\frac{1}{k^{\beta_2}})$, with $\beta_1 \geq 1$, $\beta_2 \geq 1$, with replication factor $d=2$, and arrival rate, $\lambda > \frac{1}{k}$,  the mean file access delay,
$$
 \bar{W}_s(n,k) \leq \Phi_1(k,\lambda,b,\tau)+M(k)
 $$
 $$  \mbox{if}, \,\,\,  2b^2 \log k \geq \tau^2 (\log 4\log k - \log \log (k/\lambda)-1)
 $$ 
 $$
 \mbox{otherwise} \,\,\, \bar{W}_s(n,k) \leq  \Phi_2(k,\lambda,b,\tau) + M(k),\,\,\mbox{where},
 \vspace{-3mm}
 $$
 \vspace{-3mm}
 \begin{eqnarray}
  \vspace{-3mm}
 && \Phi_1(k,\lambda,b,\tau)=2b\log k + \frac{\log 4\log k - \log \log (k/\lambda)-1}{k} \nonumber \\
 && + \frac{2}{k^4}\frac{\log (k/\lambda)}{(4\log k)}, \,\,\,\,\, M(k)=\mathbb{E} [ \displaystyle \max_{i=1,2,\ldots,k}R_i],\,\, \mbox{and} \nonumber
 \end{eqnarray}
 \vspace{-4mm}
 \begin{eqnarray}
&& \Phi_2(k,\lambda,b,\tau)=\tau \sqrt{2 \log k}\sqrt{\log 4\log k - \log \log (k/\lambda)-1} \nonumber \\
 && + \frac{\log 4\log k-\log \log (k/\lambda)-1}{k} +\frac{2\log(k/\lambda)}{4k^4  \log k} \nonumber
 \end{eqnarray}
\end{theorem}

\begin{remark}
We derive a sub-exponential maximal inequality and carefully choose system parameters using double exponential decay property for the proof. The details can be found in Appendix~\ref{appendix:int}.
\end{remark}

\begin{remark}
 For simplicity of calculation, we assume that $X_c^{(j,i)}$ is sub-exponential with $(\tau^2,b)$ satisfying, $\tau=\mathcal{O}(\frac{1}{k^{\beta_1}}), b= \mathcal{O}(\frac{1}{k^{\beta_2}})$, with $\beta_1 \geq 1$, $\beta_2 \geq 1$. However, this is not a strict requirement. Given any dependence of $\tau$ and $b$ with $k$, we can always choose appropriate problem parameters, and hence obtain an upper bound on mean delay. However, in most cases (example, exponential distribution, shift plus exponential distribution) this condition holds.
\end{remark}

\vspace{-1.5mm}
\subsection{Statistics of Residual Time}
 In case of general service distribution, we need to understand the statistics of the residual time. We refer to \cite{bramson_sampling}, where it is shown how the queuing dynamics of a large data center can be decomposed to the queuing behavior of an M/G/1 queue. For an M/G/1 queue with poisson arrival rate $\lambda_1$ and service distribution $X_1$, one can compute the mean residual time via standard renewal theory arguments. The expected  residual time of such a queue is given by, $\mathbb{E}(R) = \frac{1}{2}\lambda_1 \mathbb{E}(X_1^2)\frac{1}{\rho}$ (refer \cite{adan_queueing}), where $\rho:=\frac{\lambda}{k}$ is the utility factor of the queue. One can readily check the validity of the result by plugging in the exponential distribution. If $X_1 \sim \exp(k)$, $\mathbb{E}(R)=\frac{1}{k}$, which is consistent since $R$ and $X_1$ are identically distributed.  We see that, in order to compute $M(k)$, the first moment information is not sufficient. We can also compute the higher order moments of $R$ from the moments of $X_1$ in the following way ( \cite{adan_queueing}): $\mathbb{E}(R^n)=\frac{\lambda_1 \mathbb{E}(X_1^{n+1})}{n+1}\frac{1}{\rho}$. One can also compute the moment generating function of $R$, $\tilde{R}(s) = \frac{\lambda(1-\tilde{X}_1(s))}{s}\frac{1}{\rho}$ (\cite{adan_queueing}), where $\tilde{X}_1(s)$ is the moment generating function of $X_1$. 

 We now extend our analysis from M/G/1 queue to a system of queues in data centers. For that, we now need to compute the effective arrival rate under the corresponding scheduling policy it is operating on. If we employ the scheduling policy of Section~\ref{sec:system_model} in a data center scenario, and from \cite{bramson_sampling}, we know that the effective arrival rate in each of the selected queue with queue length, $\hat{Q}_i(n,k)=m$ will be, 
$\lambda_m := \frac{\lambda [\mathbb{P}(Y \geq m)^d- \mathbb{P}(Y \geq m+1)^d]}{\mathbb{P}(Y=k)}$, ( $Y$ has the distribution identical to the queue lengths), which is state dependent.  Under this scheme, the mean of the residual time,
\vspace{-2mm}
\begin{equation}
\mathbb{E}(R_i) = \mathbb{E}[\mathbb{E}(R_i|\hat{Q}_i(n,k))]= \frac{1}{2}\lambda \mathbb{E}(X_c^i)^2 \frac{1}{\rho}\nonumber
\vspace{-1mm}
\end{equation} 
where the last equality follows from a telescoping sum. Similarly, the higher order statistics and moment generating function can also be simplified, using iterative expectation and telescoping sums (using the definition of $\lambda_m$). 
\vspace{1mm}
\subsubsection{An Upper Bound on $M(k)$}
We can compute an upper bound on $M(k)$ as a function of the moment generating function of the service time $\tilde{X}_1(.)$. 

\begin{lemma}
$M(k) \leq \displaystyle \min_{s} \frac{1}{s}\log  ( \frac{ k^2 (1-\tilde{X}_1(s))}{s} )$
\end{lemma}
\vspace{1mm}
 In simulations (Section~\ref{sec:numerical_high}), instead of computing the exact $M(k)$, we use the above upper bound.
\vspace{-1mm}
\subsection{(Special Case): Memory-less Distribution}

For the special case of Exponential distribution (an exponential distribution with mean $\frac{1}{k}$, is $(\frac{1}{k^2}, \frac{1}{k})$ sub-exponential), we have the following result:
\begin{theorem}
\label{thm:exp_result_2}
Under the large system limit ($L \rightarrow \infty$), with $X_r \sim \exp(1)$,  and $X_c^i \sim \exp(k)$, ( $i=\lbrace 1,2,\ldots,k \rbrace$) i.i.d, the mean file access delay is given by,
\begin{eqnarray}
&& \bar{W}_s(n,k) \leq \frac{2 \log k}{k}+ \frac{\log 4\log k}{k} - \frac{\log \log (k/\lambda)}{k}  \nonumber \\
&& + \frac{2\log(k/\lambda)}{4 k^4 \log k}:=\Phi_3(k,\lambda) \,\,\, \mbox{if,} \,\,\,  2\log k \geq \log4\log k \nonumber \\ 
&& -\log \log (k/\lambda), \,\,\,  \,\,\, \mbox{otherwise,} \,\,\, \nonumber \\
&& \bar{W}_s(n,k) \leq  \frac{\sqrt{2 \log k}}{k} \sqrt{\log 4\log k - \log \log (k/\lambda)} \nonumber \\
&&  + \frac{\log 4\log k}{k} - \frac{\log \log (k/\lambda)}{k}+ \frac{2\log(k/\lambda)}{4 k^4\log k}:=\Phi_4(k,\lambda)  \nonumber
\end{eqnarray}
\end{theorem}

We observe that, the last term in both the above expression decays as $\frac{1}{k^4}$. Also, if we are working with a large enough $k$, the condition, $2\log k \geq  \log 4\log k  -\log \log (k/\lambda)$ will be satisfied, and in the first expression, the dominating term will be the first term. In that regime, 
\vspace{-2mm}
\begin{equation}
\label{eqn:exp_mean_delay}
\bar{W}_s(n,k) = \mathcal{O}(\frac{\log k}{k})\approx \mathcal{O}(\frac{H(k)}{k})
\vspace{-2mm}
\end{equation}
 where $H(k)$ is the harmonic number (i.e., $H(k)=\sum_{i=1}^{k}\frac{1}{i}$).
 
 \vspace{-1.5mm}
 \subsection{Gain over a Naive Replication Based System}

In the naive replication based system, we analyze $(d,1)$ scheme. The load balancing scheduling policy dictates that, the scheduler forwards the read request to the server (among $d$ servers containing the file) which has the least queue length. Hence, this scenario is identical to the one where the schedulaer choses $d$ queues at random and joins the one having least queue length (refer to Figure~\ref{fig:rep_vs_era}). We are interested in the quantity, $\bar{W}(d,1)$. Therefore, we can apply the result of \cite{bramson_sampling} directly in this case, (as $d \geq 2$) for queue length statistics.

The gain over a naive replicated system is defined as: $G = \bar{W}(d,1)-\bar{W}_s(n=dk,k)$. Therefore, the upper bounds on $\bar{W}_s(n=dk,k)$ will infer lower bounds on $G$.

\subsection{Analysis of Tail Latency for memory less servers}
 In this section, we provide an analysis of the tail bound on latency for servers with exponential service time. Let $W_s(n,k):= \max_{1 \leq i \leq k} \sum_{j=1}^{\hat{Q}_i(n,k)} X_c^{(j,i)}$ denote the latency of a split-and replicated data center, we have:
 
 \begin{theorem}
 \label{thm:tail_latency}
Consider the servers have exponential service time with $(\tau^2= 1 /k^2, b=1/k)$. We have,
\vspace{-2mm}
\begin{eqnarray}
\mathbb{P} ( W_s(n,k) > t  ) \leq k \exp \lbrace - \frac{k^2}{2 r } (t-\frac{r}{k})^2  \rbrace + \epsilon, \,\,\,  \mbox{if} \,\, 0 \leq t \leq \frac{2r}{k} \nonumber 
\end{eqnarray}
\vspace{-5mm}
\begin{eqnarray}
\mathbb{P} ( W_s(n,k) > t ) \leq k \exp \lbrace - \frac{k}{2} (t-\frac{r}{k})  \rbrace + \epsilon \,\,\,  \mbox{if} \,\, t \geq \frac{2r}{k} \nonumber \vspace{-2mm}
\end{eqnarray}

where, $r = \log ( \frac{\log \epsilon/k}{\log \lambda/k} )$, and $\epsilon >0$, a small real number.
 \end{theorem}

\section{Latency of an Erasure Coded Data Center}
\label{sec:coded_system}
We now analyze the latency of an erasure coded data center with a huge number of servers. We will consider the following $3$ settings: a) latency with low arrival rate b) latency of a high redundancy erasure code ($n=dk, d \in \mathbb{Z}_+$ with positive arrival rate, and c) latency of a low redundancy coded system ($ 1 <d <2$) with positive arrival rate.

\subsection{Mean Latency with Low Arrival Rate}
\label{sec:low_arrival}

In this section, we will analyze the performance of an erasure coded system with low arrival rate. Note that the analysis in this section will hold for both high and low redundancy erasure codes.

 If the arrival rate is low, we can assume the queues are empty at the time of a file-access request. The scheduling policy (load-balancing)  works as follows: since all the queues are empty, the file-access request is forwarded to a set of $k$ servers chosen uniformly at random out of $n$ servers. In low arrival scenario, we can expect that all the queues start serving the request instantly. The mean file access delay is given by,
 \vspace{-1mm}
\begin{equation}
\bar{W}(n,k)=\mathbb{E} [ \max_{i=1,2,\ldots,k} X_c^i ] \label{eqn:zero_traffic} 
\vspace{-2mm}
\end{equation} 

where, $X_c^i$, $(i=1,\ldots,k)$ is the service time for the $i$-th server. Note that, Equation~\ref{eqn:zero_traffic} is independent of the redundancy factor $d$ being an integer or a rational number, and hence it is valid for both high and low redundancy codes.

 We now consider high redundancy erasure codes ($n=dk$, $d$ integer) and compare the mean file delay with $(d,1)$ (naive replication with replication factor $d$) system. For erasure coded system, each file is chunked into $k$ parts and hence to match the storage requirements, one needs to compare $d$ replicated system to an $(n,k)$ erasure coded system with $n=dk$. 

 Since the queue lengths are almost zero, the mean file access delay for the naive replication based system will be, $\bar{W}(d,1)=\mathbb{E}(X_r)$, where $X_r$ denotes service time of a single server. We have,
 \vspace{-2mm}
\begin{equation}
G=\bar{W}(d,1) - \bar{W}(n,k)= \mathbb{E}(X_r) - \mathbb{E} [ \max_{i=1,2,\ldots,k} X_c^i  ] \label{eqn:diff_low}
\vspace{-2mm}
\end{equation}
\vspace{-2.5mm}

A positive gain implies that erasure codes reduce mean latency. We compute the exact gain for two typical distributions: a) Exponential and b) Shift plus Exponential. In case of exponential distribution, since each file is chunked into $k$ parts, we can expect $X_c^i\sim \exp (k)$ and for naive replicated system, $X_r\sim \exp(1)$, so that $\mathbb{E}(X_c^i)=\frac{1}{k}\times\mathbb{E}(X_r)$. For exponential distribution, we can compute the expected $k$-th order statistic in closed form solution (\cite{order_stat_exp}),$
\mathbb{E} [ \max_{i=1,2,\ldots,k} X_c^i  ] =\frac{H(k)}{k} $
where, $H(k)$ is the $k$-th harmonic number. The gain is, $G=1-\frac{H(k)}{k}$, and $G>0$ (since, $\frac{H(k)}{k}<1$).

In case of shift plus exponential distribution, $X_r \sim c+\exp(\frac{1}{1-c})$, so that $\mathbb{E}(X_r) = 1$. Similarly, $X_c^i = \frac{c}{k} + \exp(\frac{k}{1-c})$. Therefore, $
\mathbb{E} [ \max_{i=1,2,\ldots,k} X_c^i ] = \frac{c}{k}+\frac{H(k)(1-c)}{k} $
 The gain is $G=1-\frac{H(k)}{k} +\frac{c}{k}(H(k)-1)$, and $G >0$.
 
 \subsection{Mean Delay with Low Arrival and Redundant Requests}
 \label{sec:redun_req_zero_arrival}
 
 Following the setup in Section~\ref{sec:low_arrival}, we now allow the load scheduling policy to send the request to $k+\Delta$ servers instead of $k$ servers, where $\Delta$ can be seen as the amount of redundancy. This scheme also allows the system to be resilient to at most $\Delta$ straggling servers. In \cite{lee_distributed}, the authors showed that straggling of a few servers can severely harm the overall performance of a system, and hence a practical scheduling policy should be robust to stragglers. Since we are working in a low arrival region, the scheduler chooses $k+\Delta$ servers out of $n$ servers uniformly at random. With this scheduling, the mean file access delay is given by, $
 \bar{W}_R (n,k) = \mathbb{E} [  (X_c^i)_{(k:k+\Delta)} ] $
 where $(X_c^i)_{(k:k+\Delta)}$ is the $k$-th order statistic (smallest) among $k+\Delta$. We now show that in low arrival regime, it is always advantageous to allow redundant requests.
 \vspace{1mm}
 \begin{lemma}
 \label{lem:low_arrival_redundancy}
 $\bar{W}_R (n,k) \leq \bar{W}(n,k)$.
 \end{lemma}
 
 The proof is deferred to the Appendix.
 
 \begin{remark}
 Lemma~\ref{lem:low_arrival_redundancy} shows that redundant request yields better performance in terms of mean delay with low arrival rate. We now consider a special case of exponential distribution. From Section~\ref{sec:low_arrival}, $\bar{W}(n,k)=\frac{H(k)}{k}$. From the order statistics of exponential distribution (\cite{order_stat_exp}), $\bar{W}_r(n,k)= \frac{H(k+\Delta)-H(\Delta)}{k}$. Notice that, both $\bar{W}_R(n,k)$ and  $\bar{W}(n,k)$ have $k$ terms in the numerator, and since $H(.)$ of harmonic number, we conclude, $\bar{W}_R (n,k) \leq \bar{W}(n,k)$.
 \end{remark}
 
 \section{Mean Delay Analysis with high Redundancy}
 \label{sec:code_high_red}
 
We now analyze a high redundancy erasure coded system, i.e., $(n,k)$ codes with $n=dk$, $d\in \mathbb{Z}_+, d \geq 2 $. The analysis is done assuming that the file request rate is not low, i.e., the queuing effects can not be ignored.
 \vspace{-3mm}
 \subsection{$k$-split scheduling policy}
 In this setting, we fork the input request to $k$ least loaded servers among $n=dk$ servers. Here, instead of analyzing the performance of this original system, we analyze the performance of a relatively worse system. We divide the $dk$ servers into $k$ batch of $d$ servers each and read $1$  out of each batch (the queue in the batch with the smallest queue length) . We claim that the expected delay of this scheduling policy will be an upper bound on the original load balancing one (smallest $k$ out of total $dk$ servers). This is because we are selecting one queue from each batch, and it may be possible the $k$ queues being read using this policy are not the smallest $k$ queues, since a batch of $d$ queues may have more than $1$ queue that falls among the smallest $k$ queues . 
 
 Since the files are MDS coded, any $k$ out of $n$ chunks are sufficient to recover the entire file. Hence, unlike in the split-and-replication based system, it does not matter how the partition of $dk$ file chunks in $k$ batches of $d$ files each is done. In particular, $\bar{W}(n,k) \leq \inf_{p \in \mathcal{P}} \bar{W}_p(n,k) \leq \bar{W}_{\tilde{p}}(n,k)$, where $\tilde{p}$ is a particular partition, and $\mathcal{P}$ is a set of partitions. Now, given the system model and scheduling policy, the mean latency with partition $\tilde{p}$ is identical to that of a split-and-replicate system, $\bar{W}_s(n,k)$, and thus, $\bar{W}(n,k) \leq \bar{W}_s(n,k)$. Therefore all the upper bounds on mean and tail latency proved in Section~\ref{sec:method} can be extended to this setting.
 \vspace{1mm}
 \begin{remark}
 Tailoring to the special case of exponential service times, from Equation~\ref{eqn:exp_mean_delay}, $\bar{W}(n,k)=\mathcal{O}(\frac{H(k)}{k})$, which is precisely the result of \cite{srikant_mean} (up to a constant factor). The sub-optimality of our result roots from 2 reasons: 1) we provide a generic scheme for all sub-exponential (which is a very rich class) service times and 2) instead of analyzing the coded system, we analyze split-and-replicate system whose performance is worse. Also, the gap between a coded and split-and-replicate system is reasonably small, and this is validated numerically in Section~\ref{sec:numerical_high}.
 \end{remark}
 
 \vspace{-1mm}
 \section{Mean Delay Analysis with low Redundancy}
\label{sec:batch_sampling}
 Upto now, we analyze cloud storage systems with integer replication factor. In practical applications, even a redundancy factor of $2$ is often prohibitive owing to storage constraints. For example, an industry standard MDS code (used by Facebook data centers, \cite{rashmi_thesis}) has a redundancy of $\frac{14}{10}=1.4$. Hence, we need to analyze erasure coded systems with low redundancy.
 \vspace{-1mm}
\subsection{System Model}
We want to analyze $\bar{W}(n,k)$, where, $ 1 < \frac{n}{k} < 2$. For this, we employ the scheduling policy namely ``batch sampling'' (\cite{batch_srikant}). A file-access request is equivalent to downloading $k$ distinct chunks of the file stored in different servers using an $(n,k)$ MDS code. Hence, a file-access request can be thought as a batch or job consisting of $k$ tasks. 

We assume that the data center has $L$ servers, the batch arrival is a Poisson process with rate $\lambda \frac{L}{k}$. Note that the arrival rate here is $\frac{1}{k}$-th of the arrival rate we considered in Sections~\ref{sec:low_arrival} and \ref{sec:method}. The service time of all the servers are independent and identically distributed as exponential distribution with mean $1$. We also want the batch sizes not to be very small, and hence we assume $k=\Theta(\log L)$.
\vspace{-1.5mm}
\subsection{Batch Sampling}
 ``Batch sampling'' works as follows: when a file-access request for a particular file (say file $i$), consisting of $k$ tasks arrive, the scheduler first probes (samples) the $n$ servers containing file $i$. Now, out of this chosen $n$ servers, the $k$ tasks are forwarded to least loaded (in terms of queue length) $k$ distinct servers, one for each server. Since all files are being requested uniformly at random, one can view the system as if an arriving batch of $k$ tasks choose randomly a set of $n$ servers. Since the request is forwarded to $k$ distinct servers, we would be able to recover file $i$, as it is encoded via an MDS code.

Under the batch sampling scheme the steady state queue length distribution under large system limit is given in \cite{batch_srikant}. Define $d=\frac{n}{k}$ as the probe ratio and let $\pi_i$ denote the probability that in steady state the server has a queue length $i$. For $1<d<2$, we have,
 \[
    \pi_i=\left\{
                \begin{array}{ll}
                  1-\lambda, \,\,\, \,\,\, i=0,\\
                  (1-\lambda)\lambda^i d^i, \,\,\, 1\leq i \leq \bar{Q}_{max} -1\\
                  1-(1-\lambda)\frac{\lambda^id^i}{\lambda d -1}, \,\,\,\,\,\, i=\bar{Q}_{max}\\
                  0 \,\,\,\,\,\,\,\,\, \mbox{else, where} \,\,\, \bar{Q}_{max} = \left \lceil{\frac{\log \frac{d-1}{d(1-\lambda)}}{\log (\lambda d)} } \right \rceil
                \end{array}
              \right.
  \]

\subsection{Mean Delay Analysis under Batch Sampling}
\label{subsec:delay_analysis_non_integer}

We will use the notation of Section~\ref{sec:method}. For an $(n,k)$ MDS code, with batch sampling, assume that the $k$ least loaded servers have queue lengths, $\hat{Q}_i(n,k), i=\lbrace 1,2,\ldots,k\rbrace $ . A reading (download) will complete when all $k$ queues are read. So, the mean delay is,
 $\bar{W}(n,k)= \mathbb{E}[\max_{i=1,2,\ldots,k}\sum_{j=1}^{\hat{Q}_i(n,k)+1} X_c^{(j,i)}]
$ where $X_c^{(j,i)}$ are iid, exponentially distributed with mean $1$. Also define $\hat{Q}_{(l)}(n,k)$ as the $l$-th smallest order statistic of steady state queue length, i.e., $ \hat{Q}_{(1)}(n,k) \leq \hat{Q}_{(2)}(n,k) \leq \ldots \leq \hat{Q}_{(k)}(n,k)$. The following theorem proves an upper bound on the expected latency:
\vspace{1mm}
\begin{theorem}
For an $(n,k)$ MDS code with $1<\frac{n}{k}<2$, the mean file access delay $\bar{W}(n,k)$ is characterized by,
\vspace{-2mm}
\footnotesize
\begin{equation}
\bar{W}(n,k)\leq H(k) + \sum_{l=1}^{k}\frac{1}{k-l+1} \mathbb{E}[\hat{Q}_{(l)}(n,k)] \leq H(k)+ \sum_{l=1}^{k} \mathbb{E}[\hat{Q}_{(l)}(n,k)]  \nonumber
\vspace{-2mm}
\end{equation}
\label{thm:result_2}
\end{theorem}
The proof can be found in Appendix~\ref{appendix:non_integer}.

We refer to the bound of Theorem~\ref{thm:result_2} as \textit{Bound I}. The second term in the bound on $\bar{W}(n,k)$ is a sum of expected order statistic of a set of discrete random variables, $\hat{Q}_i(n,k)$, which is often hard to characterize. One way to approximate the sum is to convert the discrete random variables $\hat{Q}_i(n,k)$ to continuous ones and work with the ordered continuous random variables. Finally, we use the results of \cite{bertsimas2006tight} to obtain a bound on the sum of continuous ordered random variables.

We first convert the discrete random variable $\hat{Q}_i(n,k)$ to continuous one $\hat{Q}_i^c(n,k)$. Note that, $\hat{Q}_i(n,k)$ is the random variable denoting  steady state queue length, which is un-ordered, and the order statistics of $\hat{Q}_i(n,k)$ is $\hat{Q}_{(l)}(n,k), 1\leq l \leq k$. The conversion from discrete to continuous random variable is done in the following way: since $\hat{Q}_i(n,k)$ is discrete, the probability mass function consists of atoms at  $\lbrace 0, 1, 2, \ldots, \bar{Q}_{max} \rbrace $. We linearly interpolate between the aforementioned points to obtain a continuous probability density function. After proper normalization, we assign this pdf to the random variable $\hat{Q}_i^c(n,k)$. 

\begin{figure}[t!]
\centering
\subfigure[]{\includegraphics[height=1.5in,width=1.715in]{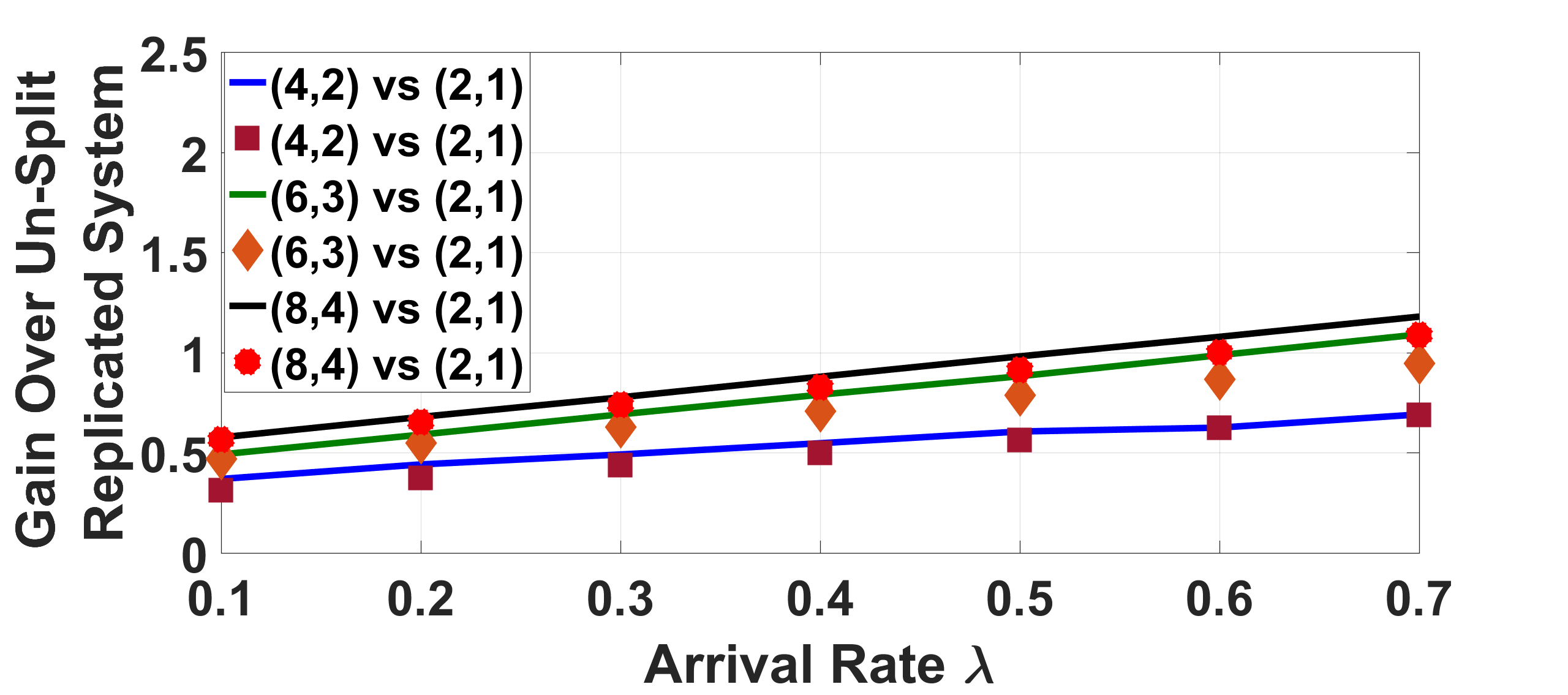}} 
\subfigure[]{\includegraphics[height=1.5in,width=1.715in]{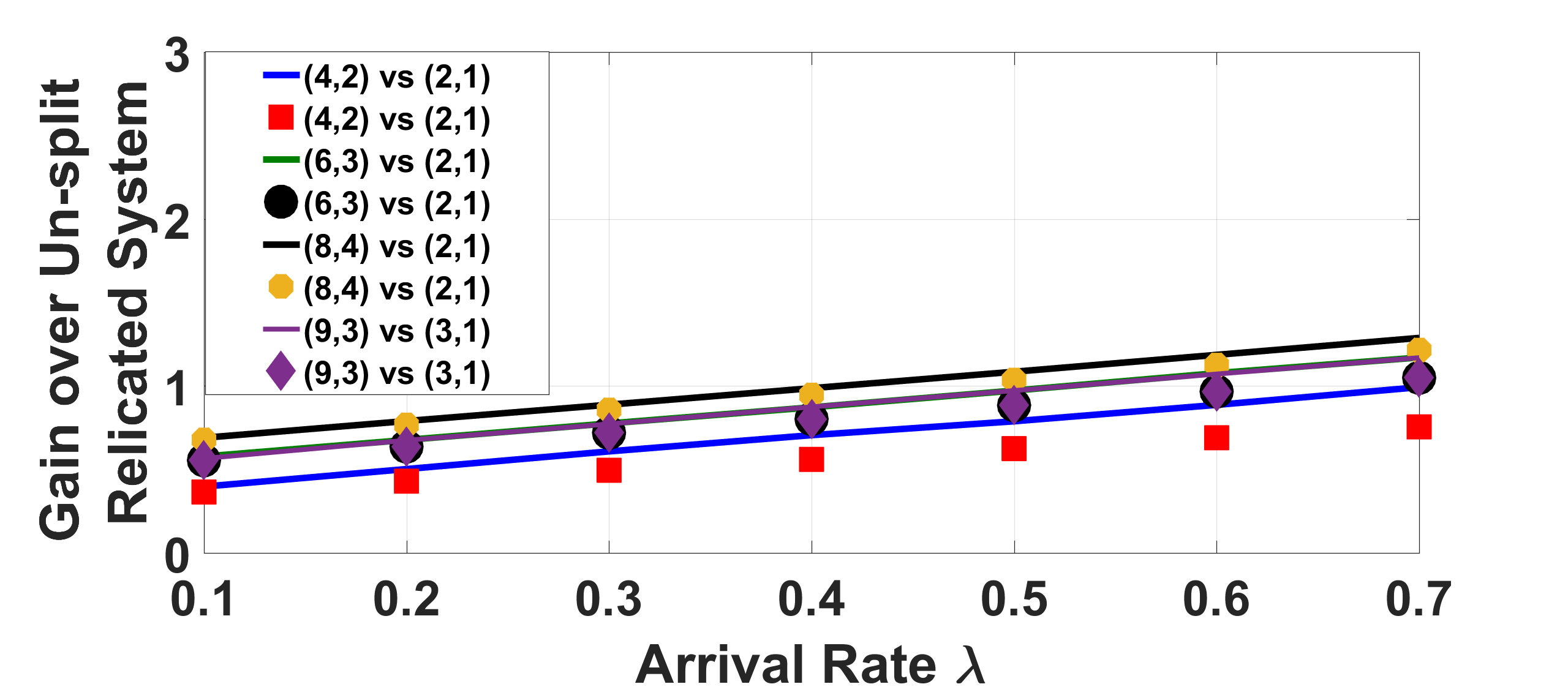}}
\vspace{-2mm}
 \caption{Relative gain $G$ is plotted for a) exponential and b) shift plus exponential with shift $0.1$. The solid lines represent data obtained from simulations, while the markers denote theoretical gain.}
\label{fig:exp_shift} 
\vspace{-3mm}
\end{figure}
 
 Under the mean field limit, the queue lengths of the servers at steady state, $\hat{Q}_i(n,k)$, are independent and identically distributed, and hence after linear interpolation, $\hat{Q}^{c}_i(n,k)$, are iid continuous random variables supported in $[0,Q_{max}]$. \cite{bertsimas2006tight} provides tight upper-bounds on sum of expected order statistics of continuous random variables with a given mean (let $\mu_2$) and variance (let $\sigma_2^2$). From the interpolation of the pdf of $\hat{Q}_i^{c}(n,k)$, one can numerically compute $\mu_2$ and $\sigma_2^2$ and upper bound the sum of $k$ order statistics of $\hat{Q}^{c}_{(l)}(n,k)$ :
 \vspace{-4mm}
 \begin{eqnarray}
   \sum_{l=1}^{k} \mathbb{E}[\hat{Q}^{c}_{(l)}(n,k)]  \leq   \min_z  ( kz+ \frac{k}{2}(\mu_2 -z+ 
   \sqrt{(\mu_2 - z)^2+ \sigma_2^2}) ) \nonumber  
   \vspace{-3mm}
 \end{eqnarray}
 We will use this as an approximation to $\sum_{l=1}^{k}\mathbb{E}[\hat{Q}_{(l)}(n,k)]$. Combining with Equation~\ref{eqn:delay_non_integer} and using the fact that $\frac{1}{k-l+1} \leq 1$ for $l=1,2,\ldots,k$, we have a bound on expected delay under low redundant erasure codes,
 \vspace{-3mm}
\begin{eqnarray}
\bar{W}(n,k)  \lessapprox  H(k)+k\min_z ( z+ \frac{1}{2}(\mu_2 -z  +\sqrt{(\mu_2 - z)^2+ \sigma_2^2}) ) \nonumber
\end{eqnarray}
We refer to this as \textit{Bound II}.

\section{Simulations}
\label{sec:simulations}
In this section, we will provide numerical evaluations of the mean file access delay and compare the tightness of the bounds obtained in Sections~\ref{sec:code_high_red} and \ref{sec:batch_sampling}. 
\subsection{High Redundancy Erasure Coded Systems}
\label{sec:numerical_high}

We are interested in the relative gain of the $(n,k)$ erasure coded system ($n=dk$) to that of a $(d,1)$ naive replicated system.  The gain (as explained in Section \ref{sec:low_arrival}),
\vspace{-2mm}
$$
G = \bar{W}(d,1)-\bar{W}(n,k)=\mathbb{E} (\sum_{j=1}^{\hat{Q}}X_r^j)-\bar{W}(n,k)
\vspace{-1mm}
$$
where $X_r^j$ is the service distribution (with mean $1$, see \ref{sec:system_model}), and $\hat{Q}$ is the queue-length under the scheduling policy of \ref{sec:system_model}. We know the queue-length distribution of $\hat{Q}$ (which is double exponential), and so the first term can be computed via Monte-Carlo techniques. In Section~\ref{sec:code_high_red}, we obtain  upper-bounds $\bar{W}(n,k)$, and therefore we can compute a lower bound on the gain $G$. We denote this lower bound as the theoretical gain, and want to compare it to the numerically observed gain.

 We evaluate the mean file delay of both erasure coded and naive-replication based system under large system limit. In particular we compared the gain of $(4,2)$, $(6,3)$ and $(8,4)$ MDS code to a $(2,1)$ naive replication code and $(9,3)$ MDS code to a $(3,1)$ naive replication code. We assume the arrivals form a Poisson process, and the relative gain $G$, is computed for different arrival rate $\lambda$. We assume the following service time distributions: i) Exponential ii) Shift plus exponential and iii) Weibull distribution. Our choice is motivated by \cite{lee_distributed} and \cite{rashmi_thesis} where it is  shown that shift plus exponential and heavy-tailed distributions capture the real behavior of data servers.
 
 In Figure~\ref{fig:exp_shift}, the gain $G$ is plotted with various arrival rate $\lambda$.  The solid lines represent data obtained from simulations, while the markers denote theoretical gain. For Figure~\ref{fig:exp_shift} (a), servers are exponentially distributed with $X_r\sim \exp(1)$ and $X_c^i \sim \exp(k)$ for $i=1,\ldots,k$. Similarly for From Figure~\ref{fig:exp_shift} (b) we use shift plus exponential distribution with shift, $c=0.1$. $X_r = c+Y_r$, where $Y_r \sim \exp(\frac{1}{1-c})$ and $X_c^i = \frac{c}{k}+ Y_c^i$ with  $Y_c^i\sim \exp(\frac{k}{1-c})$ ( $c<1$), so that $\mathbb{E}(X_c^i)=\frac{1}{k}$. Refer to Appendix~\ref{appendix:numerical} for a sub-exponential characterization of the distribution. From Figure~\ref{fig:exp_shift}, we see that the $k$-split model closely approximates the original system. 
 
\begin{figure}[t!]
    \centering
    \includegraphics[width=0.45\textwidth]{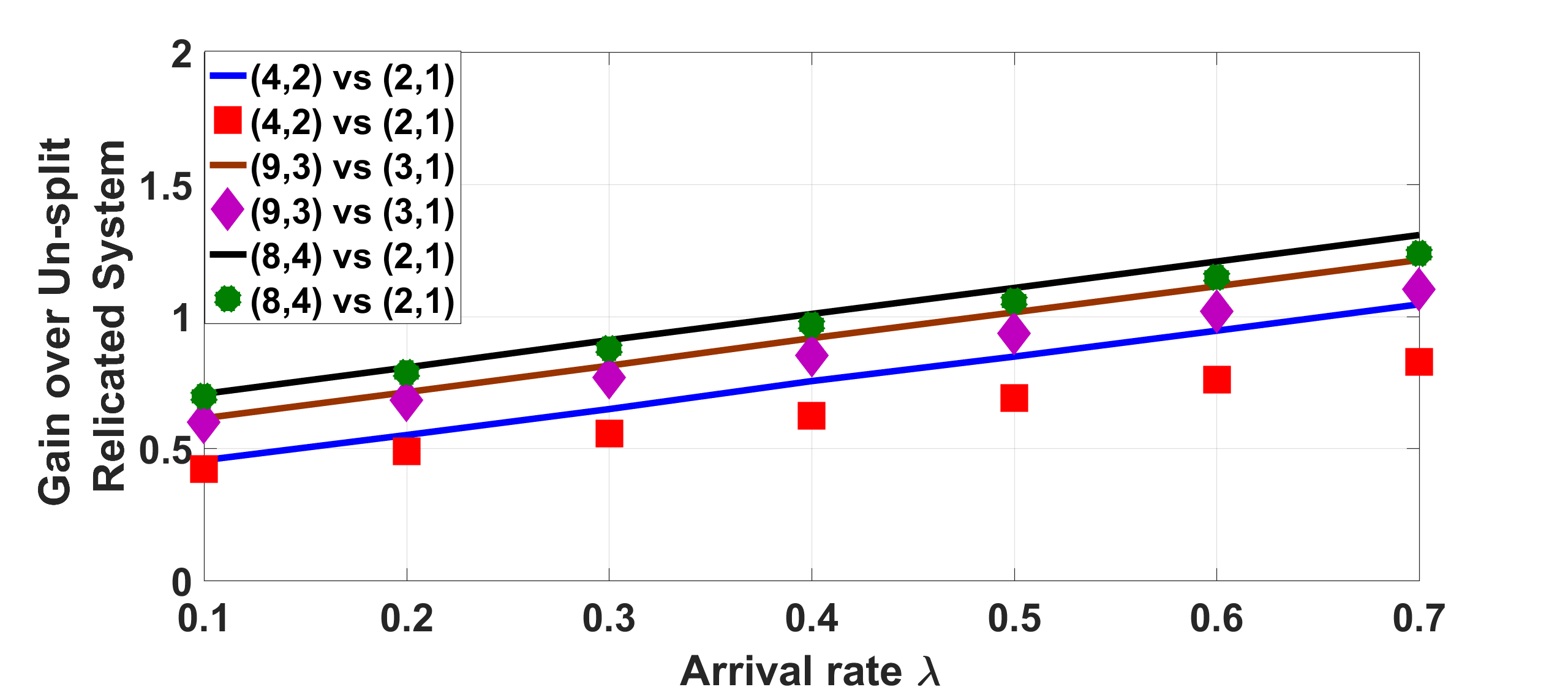}
    \caption{Relative gain $G$ with different arrival rate for ``heavy tailed'' Weibull distribution with parameter as in Section~\ref{sec:numerical_high}. The solid lines represent data obtained from simulations, while the markers denote theoretical gain
     }
    \label{fig:wei}
\end{figure}

 We next consider a typical ``heavy tail'' distribution, namely the Weibull distribution, and the density function is given by, $f(x;m)=m x^{(m-1)} e^{-x^m}$, with $m >0$ (supported on $[0,\infty]$).  We have, $X_c^{(j,i)}  \sim Wei(m)$ such that, $\frac{1}{m}\Gamma(\frac{1}{m})=\frac{1}{k}$.  This ensures, $\mathbb{E}(X_c^{(j,i)})=\frac{1}{k}$, since $\mathbb{E}(X_c^{(j,i)})=\Gamma(1+\frac{1}{m})=\frac{1}{m}\Gamma(\frac{1}{m})$. Weibull distribution trivializes to an exponential distribution for $m=1$.  Here $\Gamma(.)$ denotes gamma function. In Appendix~\ref{appendix:numerical} we characterize the sub-gaussian parameters of Weibull distribution. The relative gain  is shown in Figure~\ref{fig:wei}.
 
 \subsection{Low Redundancy Erasure Codes}
\begin{figure}[t!]
\centering
\subfigure[]{\includegraphics[height=1.5in,width=1.7in]{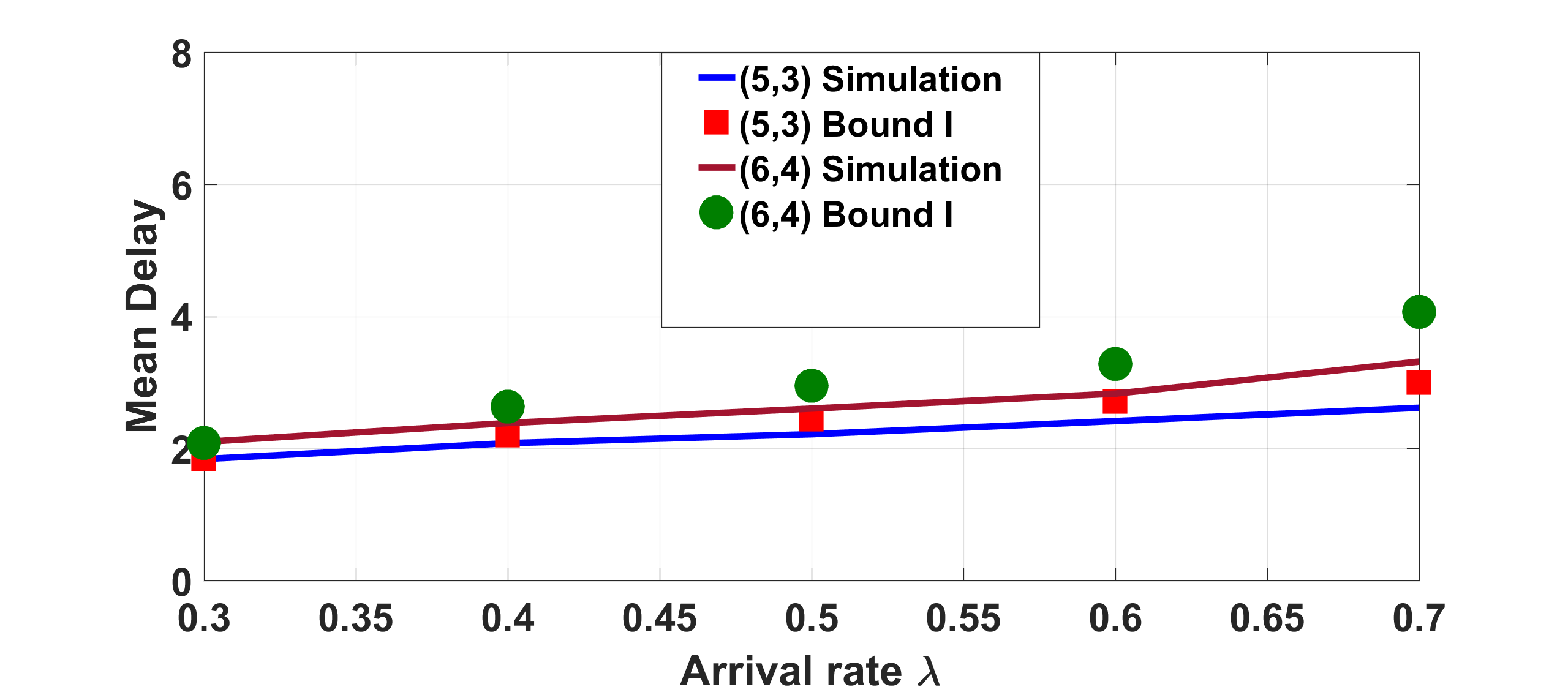}} 
\subfigure[]{\includegraphics[height=1.5in,width=1.7in]{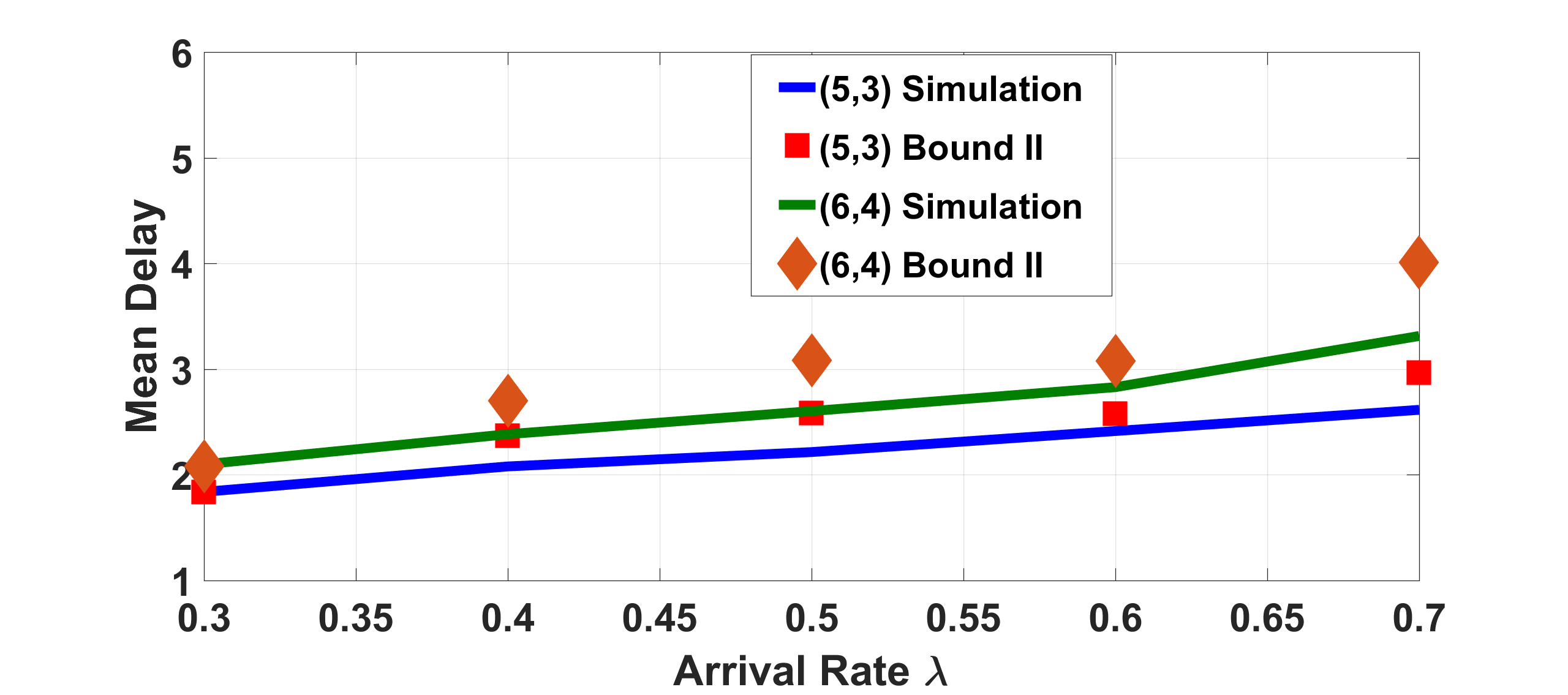}}
\vspace{-2mm}
 \caption{Mean file access delay and Bound I and II (from Section~\ref{sec:batch_sampling} for low redundant erasure coded system. The solid lines represent data obtained from simulations, while the markers denote theoretical bound.}
\label{fig:non_int_red} 
\vspace{-3mm}
\end{figure}

 We now analyze the mean file access delay for an $(n,k)$ MDS code with $1<\frac{n}{k}<2$. We compare the average file access delay to the bounds given in Section~\ref{sec:batch_sampling}. As explained in Section~\ref{sec:batch_sampling}, we assume that the service distribution is exponential with unity mean and independent across servers. Figure~\ref{fig:non_int_red} shows the variation of the relative gain with respect to $\lambda$. We analyze for a relatively higher values of $\lambda$ to avoid the triviality of $\bar{Q}_{max}$ going to $0$. Figure~\ref{fig:non_int_red} (a) shows the tightness of Bound I (refer Theorem~\ref{thm:result_2}) and Figure~\ref{fig:non_int_red}(b) shows the the error in approximation of Bound II. We see that Bound II is relatively loose with respect to Bound I because of the discrete to continuous approximation.
 
\section{Conclusion and Future Work}
We analyze a coded system with both high and low redundancy for general sub-exponential service time distribution. We obtain a lower bound on gain, and analyze the lower-bound for exponential, shift plus exponential distribution and weibull distribution. Our future plan is to incorporate redundant request with this framework. For low redundant codes, we restrict ourselves to exponential service time distribution. One immediate future work would be to analyze a low redundancy erasure coded system for general service time distribution. Also, it would be interesting to study the effect of flooding with non-zero cancellation cost with general service time distribution. We keep these research problems as our future endeavors.

\bibliographystyle{IEEEtran}
\bibliography{IEEEabrv,avishek-techreport}
\clearpage

\section*{APPENDIX}
\section{Split-and-Replicate System} 
\label{appendix:int}

\vspace{1mm}
\begin{lemma}
$M(k) \leq \displaystyle \min_{s} \frac{1}{s}\log  ( \frac{ k^2 (1-\tilde{X}_1(s))}{s} )$
\end{lemma}
\vspace{2mm}

\begin{IEEEproof}
Using Jensen's inequality (for convex function $f(x) = \exp(x)$), for $s >0$,
\begin{eqnarray}
&&\exp(s\mathbb{E}\displaystyle \max_{i=1,\ldots,k} R_i  ) \leq  \mathbb{E}\exp (s \max_{i=1,\ldots,k} R_i ) \nonumber\\
&&=\mathbb{E}\max_{i=1,\ldots,k} \exp(s R_i) \leq \sum_{i=1}^{k}\mathbb{E}\exp(s R_i) =  k \tilde{R}(s)\nonumber \\
&& =\frac{ k^2 (1-\tilde{X}_1(s))}{s} \nonumber
\end{eqnarray}
where $\tilde{X}_1(s)$ is the moment generating function of the service time evaluated at $s$. Take logarithm both sides and observe that the right hand side is a function of $s$.
\end{IEEEproof}

\begin{lemma}
 $\bar{W}_R (n,k) \leq \bar{W}(n,k)$.
 \end{lemma}
 \vspace{2mm}
 
 \begin{IEEEproof}
 Define  $Y_k = \max_{1\leq i\leq k} X_c^i$. From the definition of order statistic, $Y_k \stackrel{\text{dist.}}{=} \displaystyle \max_{S \subset [k+\Delta], |S|=k} X_S$, where $X_S$ is a collection of $k$ random variables in set $S$. In words, given a set of $k+\Delta$ random variables, if one takes any possible subset of cardinality $k$ and find the maxima, one obtains $Y_k$. On the other hand, $(X_c^i)_{(k:k+\Delta)}$ is the $k$-th smallest element chosen from $k+\Delta$ instances. From the definition of $Y_k$, it is clear that, if $S$ contains $k+1$-th, $\ldots$, $k+\Delta$-th smallest element, $Y_k \geq (X_c^i)_{(k:k+\Delta)}$, otherwise, $Y_k = (X_c^i)_{(k:k+\Delta)}$. The conclusion follows since pointwise ordering implies ordering in expectation.
 \end{IEEEproof}
 
 \vspace{1mm}
\begin{lemma}
\label{lem:sub_exp_identity}
Consider a set of $N$ i.i.d sub-exponential random variables $Z_1,\ldots,Z_N$, with sub-exponential parameter $(\tau^2,b)$. Then we have,
$$
\mathbb{E}\bigg [ \max_{1\leq i \leq N} Z_i \bigg ]  \leq \max \bigg ( \tau \sqrt{2 \log N} + \mathbb{E}Z, 2b \log N +\mathbb{E}Z \bigg )
$$
\end{lemma}

\begin{IEEEproof}
We have,
\begin{eqnarray}
&&\mathbb{E}\bigg [ \max_{1\leq i \leq N} Z_i \bigg ] = \frac{1}{\lambda} \mathbb{E} \bigg [ \log (e^{(\lambda \max_{1\leq i\leq N}Z_i)} \bigg ] \nonumber \\
&&\leq \frac{1}{\lambda} \log \bigg [ \mathbb{E} ( e^{(\lambda \max_{1\leq i\leq N}Z_i)} ) \bigg ]\nonumber \\
&& = \frac{1}{\lambda} \log \bigg [ \mathbb{E} (\max_{1 \leq i \leq N}e^{\lambda Z_i}) \bigg ] \leq \frac{1}{\lambda} \log \bigg [ \mathbb{E} \sum_{i=1}^{N} \mathbb{E}(e^\lambda Z_i) \bigg ] \nonumber \\
&&= \frac{1}{\lambda} \log \bigg [ \mathbb{E} \sum_{i=1}^{N} \mathbb{E}(e^{\lambda(Z_i - \mathbb{E}Z)} e^{\lambda \mathbb{E}Z}) \bigg ]\nonumber \\ 
&& \leq \frac{1}{\lambda}\log \bigg [ e^{(\lambda \mathbb{E}Z)} \sum_{i=1}^{N} e^{\frac{\lambda^2 \tau^2}{2}} \bigg ] \,\,\, \mbox{for} \, \lambda \leq \frac{1}{|b|} \nonumber \\
&&= \frac{1}{\lambda}\log N + \frac{\lambda \tau^2}{2}+ \mathbb{E}Z \nonumber
\end{eqnarray}

We now consider $2$ cases, regarding the choice of $\lambda$. The first two components are dependent on $\lambda$. Ideally, the optimal choice of $\lambda:=\sqrt{\frac{2 \log N}{\tau^2}}$. But there is a constraint, $\lambda \leq \frac{1}{|b|}$.
\paragraph*{Case I:}
$\sqrt{\frac{2\log N}{\tau^2}} \leq \frac{1}{|b|}$. We have,
$$
\mathbb{E}\bigg [ \max_{1\leq i \leq N} Z_i \bigg ] \leq \tau \sqrt{2 \log N} + \mathbb{E}Z
$$
\paragraph*{Case II:}
Here, $\sqrt{\frac{2 \log N}{\tau^2}} > \frac{1}{|b|} \Rightarrow \tau^2 \leq 2b^2\log N$. We have,
$$
\mathbb{E}\bigg [ \max_{1\leq i \leq N} Z_i \bigg ]  \leq 2b \log N +\mathbb{E}Z
$$

Combining two yields the result.

\end{IEEEproof}

\subsection{Proof of Theorem~\ref{thm:result_1_gen} }

We know that the distribution of $\hat{Q}_i(n,k)$ will be double exponential:
  $$
  \mathbb{P}(\hat{Q}_i(n,k) \geq r) \leq c_u (\frac{\lambda}{k})^{2^r}
  $$
  where $c_u = 1+o(1)$. For simplicity of calculation, from now on, we assume $c_u = 1$.
  
  We want to control a term of the following form:
$$
\mathbb{E} \bigg ( \max_{1\leq i \leq k} \sum_{j=1}^{\hat{Q}_i(n,k)}X_c^{(j,i)} \bigg )
$$
where, $\hat{Q}_i(n,k)$ is the queue length for the $i$-th queue, having a double exponential decay property, and $X_c^{(j,i)}$ are service times (random variables). We make the assumption that $X_c^{(j,i)}$ follows \textbf{sub-exponential} distribution. This is a valid assumption since all heavy tailed distribution like Weibull, Pareto distribution follow sub-exponential property (\cite{goldie_subexp}). Also, since all sub-gaussian random variables are sub-exponential, the class of all sub-exponential distribution is very rich. We upper bound the above expression as follows:

\begin{eqnarray}
&&\mathbb{E} \bigg ( \max_{1\leq i \leq k} \sum_{j=1}^{\hat{Q}_i(n,k)}X_c^{(j,i)} \bigg ) \nonumber \\
&=& \mathbb{E} \bigg ( \max_{1\leq i \leq k} \sum_{j=1}^{\hat{Q}_i(n,k)}X_c^{(j,i)} \mathrm{1}_{ \lbrace \max_{1\leq i \leq k} \hat{Q}_i(n,k) \leq r \rbrace } \bigg )\nonumber \\
& + & \mathbb{E} \bigg ( \max_{1\leq i \leq k} \sum_{j=1}^{\hat{Q}_i(n,k)}X_c^{(j,i)} \mathrm{1}_{ \lbrace \max_{1\leq i \leq k} \hat{Q}_i(n,k) > r \rbrace } \bigg ) \nonumber \\
&\leq & \mathbb{E} \bigg ( \max_{1\leq i \leq k} \sum_{j=1}^{r}X_c^{(j,i)} \bigg ) \mathbb{P}(\max_{1\leq i\leq k}\hat{Q}_i(n,k) \leq r)\nonumber \\
& + & \sum_{l=r+1}^{\infty} \mathbb{E} \bigg ( \max_{1\leq i\leq k} \sum_{j=1}^{l}X_c^{(j,i)} \bigg ) \mathbb{P}(\max_{1\leq i\leq k} \hat{Q}_i(n,k)=l) \nonumber \\
&\leq & \mathbb{E} \bigg ( \max_{1\leq i \leq k} \sum_{j=1}^{r}X_c^{(j,i)} \bigg ) \nonumber \\
&+ & \sum_{l=r+1}^{\infty} \mathbb{E} \bigg ( \max_{1\leq i\leq k} \sum_{j=1}^{l}X_c^{(j,i)} \bigg ) \mathbb{P}(\max_{1\leq i\leq k} \hat{Q}_i(n,k) \geq l) \nonumber 
\end{eqnarray}

where we upper bound the probability term in the first expression by $1$. We now have to control the individual terms. In order to do this, we use Lemma~\ref{lem:sub_exp_identity}

We will use this to control the first term, \\
 Define $\mathbb{E} \bigg ( \max_{1\leq i \leq k} \sum_{j=1}^{r}X_c^{(j,i)} \bigg ):= \mathbb{E}(\max_{1 \leq i \leq k} Y_i)$. Now, since $X_c^{(j,i)}$ are $(\tau^2, b)$ sub-exponential, $Y_i$ will be $(r\tau^2,b)$ sub-exponential. Using the proved maximal inequality,
\begin{eqnarray}
&& \mathbb{E}(\max_{1 \leq i \leq k} Y_i) \leq \max (\tau \sqrt{r} \sqrt{2 \log k}+ \mathbb{E}Y, 2b \log k + \mathbb{E}Y )\nonumber \\
&& = \max (\tau \sqrt{r} \sqrt{2 \log k} + \frac{r}{k}, 2b \log k + \frac{r}{k} ):= \phi(r)  \label{eqn:maximal_gen}
\end{eqnarray}

\subsubsection{Choice of $r$}
We want the following to hold: For, (small) $\delta >0$,
$$
\mathbb{P}(\max_{1 \leq i \leq k} \hat{Q}_i(n,k) >r) <\delta
$$
Without loss of generality, we assume $d=2$. We have (all the $\log$ is taken with base $2$),
\begin{eqnarray}
\mathbb{P}(\max_{1 \leq i \leq k} \hat{Q}_i(n,k) >r)  \leq k \mathbb{R}(\hat{Q}_1(n,k) \geq r+1) \leq k (\frac{\lambda}{k})^{2^{r+1}} \hspace{-3mm} :=\delta \nonumber
\end{eqnarray}

and therefore, given a choice of $\delta$, $r$ is chosen as follows,
$$
r+1=  \bigg ( \log \log (k/\delta)- \log \log (k/\lambda) \bigg ) 
$$

Putting everything together,

\begin{eqnarray}
&&\mathbb{E} \bigg ( \max_{1\leq i \leq k} \sum_{j=1}^{\hat{Q}_i(n,k)}X_c^{(j,i)} \bigg ) \leq \phi(r) + \sum_{l > r} \phi (l) k (\frac{\lambda}{k})^{2^l} \nonumber
\end{eqnarray}

Lets concentrate on the second term. From the definition of $\phi(l)$, the dependence on $l$ is linear. On the other hand, $(\frac{\lambda}{k})^{2^l}$ is doubly exponentially dependent on $l$.

Now,  we need to choose $\delta$  in such a way that, for all $l > r$,
 $$ 
 \phi (l) k (\frac{\lambda}{k})^{2^l} \leq \frac{1}{k^4} (\frac{1}{2})^l
 $$
 This is ensured as follows:
If we want, $ \phi (l) k (\frac{\lambda}{k})^{2^l} \leq \frac{1}{k^4} (\frac{1}{2})^l $, we need,
\begin{equation} 
\label{eqn:ignore_phi}
l -\log l \geq \log (4\log k) + \log \log k \phi(l) - \log \log (k/\lambda)
\end{equation}
Since, $X_c^{(j,i)}$ is sub-gaussian with mean $\frac{1}{k}$, it is not unreasonable to assume that, $\tau=\mathcal{O}(\frac{1}{k^{\beta_1}}), b= \mathcal{O}(\frac{1}{k^{\beta_2}})$, with $\beta_1 \geq 1$, $\beta_2 \geq 1$. We will show that distributions like exponential, shift plus exponential distribution satisfy this property. With this, we can ignore the $\log \log (k\phi(l)) $ term in the above expression (we can do that, since the leading terms are $l$ and $\log l$, and $\phi(l)$ has a linear dependence on $l$) . We have,
$$
 l-\log l \geq \log (4\log k)  - \log \log (k/\lambda)
 $$
If we choose $\delta$ as follows:,
$$
\delta = \frac{1}{k^3} \Rightarrow r+1 = \log (4\log k)-\log \log (k/\lambda)
$$

we have, $l -\log l \geq r+1 > r $, which implies, $l>r$, since $l\geq 1$. Also, with this choice of $\delta$, we have,
$$
r+1= \log(4\log k)-\log \log (k/\lambda)
$$
Note that, since, $\lambda > \frac{1}{k}$, we have, $r > 1 $. The case where $\lambda$ is small is dealt in Section~\ref{sec:low_arrival}.

 Plugging in, we have,

\begin{eqnarray}
&& \mathbb{E}  ( \max_{1\leq i \leq k} \sum_{j=1}^{\hat{Q}_i(n,k)}X_c^{(j,i)} )  \leq   \phi(r) + \frac{1}{k^4} \sum_{l>r} (1/2)^l \nonumber\\
& =& \phi(r) + \frac{1}{k^4} (\frac{1}{2})^r \leq  \phi (r) + \frac{1}{k^4}(\frac{1}{2})^{ \log 4\log k - \log \log (k/\lambda)-1} \nonumber \\
&=& \phi(r) + \frac{2}{k^4}\frac{\log (k/\lambda)}{(4\log k)}
\end{eqnarray}
Therefore we have the following expression,

\begin{eqnarray}
&& \mathbb{E}  ( \max_{1\leq i \leq k} \sum_{j=1}^{\hat{Q}_i(n,k)}X_c^{(j,i)} )  \leq 2b\log k + \frac{\log 4\log k}{k} - \frac{\log \log (k/\lambda)}{k}  \nonumber \\
&& \,\,\,\,\,\,\,\,\,\,\,\,-\frac{1}{k}+ \frac{2}{k^4}\frac{\log (k/\lambda)}{(4\log k)} :=\Phi_1(k,\lambda,b,\tau) \nonumber \\
&& \,\,\, \mbox{if}, \,\,\, 2b^2 \log k \geq \tau^2 (\log 4\log k - \log \log (k/\lambda)-1)  \,\,\, \mbox{otherwise,} \,\,\, \nonumber \\
&& \mathbb{E}  ( \max_{1\leq i \leq k} \sum_{j=1}^{\hat{Q}_i(n,k)}X_c^{(j,i)} )  \leq \sqrt{\log 4\log k - \log \log (k/\lambda)-1} \nonumber \\
&& \times (\tau \sqrt{2 \log k}) + \frac{\log 4\log k-\log \log (k/\lambda)-1}{k} +\frac{2\log(k/\lambda)}{4k^4  \log k}\nonumber \\
&&:=\Phi_2(k,\lambda,b,\tau) \nonumber
 \end{eqnarray}
 Hence the theorem follows.

\begin{remark}
For simplicity of calculation, we assume that the random variables $X_c^{(j,i)}$ sub-exponential with $(\tau^2,b)$ satisfying, $\tau=\mathcal{O}(\frac{1}{k^{\beta_1}}), b= \mathcal{O}(\frac{1}{k^{\beta_2}})$, with $\beta_1 \geq 1$, $\beta_2 \geq 1$. However, this is not a strict requirement. Given a dependence of $\tau$ and $b$ with respect to $k$, we can always choose appropriate $\delta$. The reason why we enforce this requirement is to make the term $\log \log k \phi(l)$ negligible in Equation~\ref{eqn:ignore_phi}. In most cases (example, exponential distribution, shift plus exponential distribution) the condition holds.
\end{remark}

\subsection{Proof of Theorem~\ref{thm:exp_result_2}}

 We need to upper bound the following term,

\begin{equation}
\bar{W}(n,k)=\mathbb{E}[ \max_{i=1,2,\ldots,k} \sum_{j=1}^{\hat{Q}_i(n,k)+1} X_c^{(j,i)}  )] \label{eqn:exact_delay_exp}
\end{equation}

Similar to the previous section, we continue like the following:

\begin{eqnarray}
&&\mathbb{E} \bigg ( \max_{1\leq i \leq k} \sum_{j=1}^{\hat{Q}_i(n,k)+1}X_c^{(j,i)} \bigg ) \nonumber \\
&=& \mathbb{E} \bigg ( \max_{1\leq i \leq k} \sum_{j=1}^{\hat{Q}_i(n,k)+1}X_c^{(j,i)} \mathrm{1}_{ \lbrace \max_{1\leq i \leq k} \hat{Q}_i(n,k) \leq r -1 \rbrace } \bigg )\nonumber \\
& + & \mathbb{E} \bigg ( \max_{1\leq i \leq k} \sum_{j=1}^{\hat{Q}_i(n,k)+1}X_c^{(j,i)} \mathrm{1}_{ \lbrace \max_{1\leq i \leq k} \hat{Q}_i(n,k) \geq  r \rbrace } \bigg ) \nonumber \\
&\leq & \mathbb{E} \bigg ( \max_{1\leq i \leq k} \sum_{j=1}^{r}X_c^{(j,i)} \bigg ) \mathbb{P}(\max_{1\leq i\leq k}\hat{Q}_i(n,k) \leq r-1)\nonumber \\
& + & \sum_{l=r}^{\infty} \mathbb{E} \bigg ( \max_{1\leq i\leq k} \sum_{j=1}^{l}X_c^{(j,i)} \bigg ) \mathbb{P}(\max_{1\leq i\leq k} \hat{Q}_i(n,k)=l) \nonumber \\
&\leq & \mathbb{E} \bigg ( \max_{1\leq i \leq k} \sum_{j=1}^{r}X_c^{(j,i)} \bigg ) \nonumber \\
&+ & \sum_{l=r}^{\infty} \mathbb{E} \bigg ( \max_{1\leq i\leq k} \sum_{j=1}^{l}X_c^{(j,i)} \bigg ) \mathbb{P}(\max_{1\leq i\leq k} \hat{Q}_i(n,k) \geq l) \nonumber 
\end{eqnarray}

  In the setup of Section~\ref{sec:general_dist},  we have, 
\begin{eqnarray}
\mathbb{E}(\max_{1 \leq i \leq k} Y_i) \leq \max(\sqrt{2r} \frac{\sqrt{\log k}}{k} + \frac{r}{k}, \frac{2 \log k}{k} + \frac{r}{k}):=\phi(r) \nonumber
\end{eqnarray}

\paragraph*{Choice of $r$}
We want the following: For, very small $\delta >0$,
$$
\mathbb{P}(\max_{1 \leq i \leq k} \hat{Q}_i(n,k) \geq r) <\delta
$$
Without loss of generality, we assume $d=2$. We have (all the $\log$ is taken with base $2$),
\begin{eqnarray}
\mathbb{P}(\max_{1 \leq i \leq k} \hat{Q}_i(n,k) \geq r)  \leq k \mathbb{R}(\hat{Q}_1(n,k) \geq r) \leq k (\frac{\lambda}{k})^{2^r} :=\delta 
\end{eqnarray}
($c_u$ is precisely $1$ for memory-less distribution)
 and therefore $\delta$ and $r$ are chosen similar to Section~\ref{sec:general_dist}.
Putting everything together,

%
%
%
%
%

 Plugging in, we have,

\begin{eqnarray}
&& \mathbb{E} \bigg ( \max_{1\leq i \leq k} \sum_{j=1}^{\hat{Q}_i(n,k)}X_c^{(j,i)} \bigg )  \leq   \phi(r) + \frac{1}{k^4} \sum_{l \geq r} (1/2)^l \nonumber\\
&=& \phi(r) + \frac{2}{k^4}\frac{\log (k/\lambda)}{(4\log k)}
\end{eqnarray}

We now calculate $\phi(r)$ with $r= \log 4\log k - \log \log (k/\lambda)$. From the definition,
$$
\phi(r) = \frac{2 \log k}{k}+ \frac{\log 4 \log k}{k} - \frac{\log \log (k/\lambda)}{k} 
$$
if, $$
2 \log k \geq \log 4 \log k -\log \log (k/\lambda)
$$
else,
\begin{eqnarray}
\phi(r) &=& \frac{\sqrt{2 \log k}}{k} \sqrt{\log 4 \log k - \log \log (k/\lambda)} + \frac{\log 4 \log k}{k} \nonumber\\
& - & \frac{\log \log (k/\lambda)}{k} \nonumber
\end{eqnarray}
Putting everything together, we obtain the theorem.

\subsection{Proof of Theorem~\ref{thm:tail_latency}}
 
  We want to find an upper-bound for the following:
 
\begin{eqnarray}
\mathbb{P} \bigg ( \max_{1 \leq i \leq k} \sum_{j=1}^{\hat{Q}_i(n,k)+1} X_c^{(j,i)} > t \bigg ) 
\end{eqnarray}

Consider the event, $\mathcal{A}= \lbrace \max_i \hat{Q}_i(n,k) \leq r -1\rbrace $. We have,
\begin{eqnarray}
\mathbb{P}  ( \max_{1 \leq i \leq k} \sum_{j=1}^{\hat{Q}_i(n,k)+1} X_c^{(j,i)} > t  )\leq \mathbb{P}  ( \max_{1 \leq i \leq k} \sum_{j=1}^{r} X_c^{(j,i)} > t  ) + \mathbb{P}(\mathcal{A}^c) \nonumber
\end{eqnarray}

We will first bound the first term.

\begin{eqnarray}
\mathbb{P} \bigg ( \max_{1 \leq i \leq k} \sum_{j=1}^{r} X_c^{(j,i)} > t \bigg ) = \mathbb{P} \bigg ( \max_{1 \leq i \leq k} Y_i > t \bigg )  \nonumber
\end{eqnarray}
where, $Y_i = \sum_{j=1}^{r} X_c^{(j,i)}$. We have,
\begin{eqnarray}
\mathbb{P} \bigg ( \max_{1 \leq i \leq k} Y_i > t \bigg ) \leq k \mathbb{P}(Y_1 >t )= k \mathbb{P}(Y_1 > \mathbb{E}Y_1 + \tilde{t})  \nonumber
\end{eqnarray}
where, $t= \mathbb{E}Y_1+\tilde{t}$. Since $Y_1$ is sub-exponential with parameters $(r \tau^2, b)=(r/k^2,1/k)$, we have,
\begin{eqnarray}
&&\mathbb{P}(Y_1 > \mathbb{E}Y + \tilde{t}) \leq e^{-\frac{\tilde{t}^2}{2 r \tau^2}} \leq \exp \lbrace -\frac{k^2}{2r} (t - r/k)^2 \rbrace \nonumber \\
 && \,\,\,  \mbox{if} \,\, 0 \leq t-\frac{r}{k} \leq \frac{ r }{k}, \,\,\, \mbox{else} \nonumber \\
&&\mathbb{P}(Y_1 > \mathbb{E}Y + \tilde{t}) \leq e^{-\frac{\tilde{t}}{2b}} \leq \exp \lbrace -\frac{k}{2}(t-\frac{r}{k}) \rbrace \nonumber
\end{eqnarray}

Combining, we get,
\begin{eqnarray}
&&\mathbb{P} \bigg ( \max_{1 \leq i \leq k} \sum_{j=1}^{r} X_c^{(j,i)} > t \bigg ) \leq k \exp \lbrace - \frac{k^2}{2 r } (t-\frac{r}{k})^2 \rbrace \nonumber \\
&& \,\,\,  \mbox{if} \,\, 0 \leq t-\frac{r}{k} \leq \frac{r }{k}, \,\,\, \mbox{else} \nonumber 
\end{eqnarray}
\begin{eqnarray}
\mathbb{P} \bigg ( \max_{1 \leq i \leq k} \sum_{j=1}^{r} X_c^{(j,i)} > t \bigg ) \leq k \exp \lbrace - \frac{k}{2} (t-\frac{r}{k}) \rbrace \nonumber 
\end{eqnarray}

Also, the second term can be made arbitrarily small as follows:
\begin{eqnarray}
\mathbb{P}(\mathcal{A}^c)= \mathbb{P}(\max_i \hat{Q}_i(n,k) \geq r) \leq k \mathbb{P}(Q_1 \geq  r) \leq k (\frac{\lambda}{k})^{2^r} \nonumber
\end{eqnarray}

\paragraph{Choice of $r$}
We want to make the probability, $\mathbb{P}(\mathcal{A}^c)$ very small, therefore:
$$
k (\frac{\lambda}{k})^{2^r} := \epsilon \Rightarrow r = \log \bigg ( \frac{\log \epsilon/k}{\log \lambda/k} \bigg )
$$

\section{low Redundant Erasure Codes}
\subsection{Proof of Theorem~\ref{thm:result_2}}
\label{appendix:non_integer}
Using ordered $\hat{Q}_i(n,k)$, $i=\lbrace 1,2,\ldots,k \rbrace$, $\bar{W}(n,k)$ can be calculated as follows (refer \cite{srikant_mean} for details):
\begin{eqnarray}
\bar{W}(n,k)&\leq & \mathbb{E}\bigg [ \sum_{j=1}^{\hat{Q}_{(1)}(n,k)+1} \max_{i=1,2,\ldots,k}X_c^{(j,i)} \bigg ] \nonumber \\
 &+&\sum_{l=2}^{k}\mathbb{E}\bigg [ \sum_{j=\hat{Q}_{(l-1)}(n,k)+2}^{\hat{Q}_{(l)}(n,k)+1} \max_{i=l,l+1,\cdot,k} X_c^{(j,i)} \bigg ] \nonumber
\end{eqnarray}

Since the queue length process, $\hat{Q}_i(n,k)$ is independent of the service random variables, $X_c^{(j,i)}$, and $X_c^i$ are iid for $i=1,2,\ldots,k$, following the proof of Proposition 3 of \cite{srikant_mean}, we get
\begin{eqnarray}
\bar{W}(n,k) &\leq & \bigg ( (1+\mathbb{E}[\hat{Q}_{(1)}(n,k)])S(k) \nonumber \\
&+&\sum_{l=2}^{k} (\mathbb{E}[\hat{Q}_{(l)}(n,k)]-\mathbb{E}[\hat{Q}_{(l-1)}(n,k)])S(k-l+1)\bigg ) \nonumber
\end{eqnarray}
After rearranging the terms in the above equation, we get
\begin{equation}
\bar{W}(n,k) \leq S(k) + \sum_{l=1}^{k} (S(k-l+1)-S(k-l))\mathbb{E}[\hat{Q}_{(l)}(n,k)] \nonumber
\end{equation}
 where, $\hat{Q}_l(n,k)$ is the steady state queue length of the corresponding queue. Since $X_c^{i}$ is exponentially distributed with mean $1$, $S(k)=H(k)$. Therefore,
 \begin{equation}
 \bar{W}(n,k) \leq  H(k)+ \sum_{l=1}^{k} \frac{1}{k-l+1}\mathbb{E}[\hat{Q}_{(l)}(n,k)]
  \label{eqn:delay_non_integer}
 \end{equation}
 Now since $k-l+1 \geq 1$ for $l=1,2,\ldots,k$, the last inequality follows.
 
 \section{Numerical Section}
 \subsection{Characterization of Sub-Exponential Parameters}
 \label{appendix:numerical}
 In this section, we will characterize the sub-exponential parameters for Shift plus exponential and Weibull-distribution. For Sub-exponential distribution, we have, $X_c^{(j,i )}= \frac{c}{k}+ Y_c$ with  $Y_c \sim \exp(\frac{k}{1-c})$ ( $c<1$), so that $\mathbb{E}(X_c^{(j,i)})=\frac{1}{k}$. Consider the first part, a constant, which is sub-exponential for any value of $\tau$ and $b=0$ (follows directly from definition of sub-exponential random variable). For simplicity, we take $\tau =1$  and $b=0$. For the second part, which is an exponential distribution, we know that it is sub-exponential with $\tau = \frac{1-c}{k}$ and $b= \frac{1-c}{k}$. So, $X_c^{(j,i )}$, which is a sum of $2$ sub-exponential random variables, is sub-exponential with $\tau^2 = 1+ (\frac{1-c}{k})^2$ and $b=\frac{1-c}{k}$.

 We now consider Weibull distribution, and the density function is given by, $f(x;m)=m x^{(m-1)} e^{-x^m}$, with $m >0$ (supported on $[0,\infty]$.  We have, $X_c^{(j,i)}  \sim Wei(m)$ such that, $\frac{1}{m}\Gamma(\frac{1}{m})=\frac{1}{k}$.  This ensures, $\mathbb{E}(X_c^{(j,i)})=\frac{1}{k}$, since $\mathbb{E}(X_c^{(j,i)})=\Gamma(1+\frac{1}{m})=\frac{1}{m}\Gamma(\frac{1}{m})$. For the sub-exponential characterization of Weibull distribution, we need a bound on the second moment. Using standard results for Weibull distribution,
 $$
 \mathbb{E}(X_c^{(j,i)})^2 = \Gamma(1+\frac{2}{m}) = \frac{2}{m}\Gamma(\frac{2}{m})=4 \frac{1}{2m}\Gamma(\frac{2}{m})
 $$
 Therefore, we can take $\tau$ and $b$ as constant multiples of the Orlicz norm $\sqrt{\frac{1}{2m}\Gamma(\frac{2}{m})}$ (refer \cite{vershynin2016high} for details). In order to choose the constants, we simulated the moment generating function of the Weibull distribution. We observe it suffices to choose $6$ as the corresponding constant.

\end{document}